\documentclass{article}

\usepackage{amsmath}
\usepackage{amssymb}
\usepackage{graphicx}
\usepackage{amsfonts}
\usepackage{amsthm}

\newtheorem{proposition}{Proposition}
\newtheorem{lemma}{Lemma}

\newcommand{\mr}{\mathrm}

\newcommand{\va}[1]{\boldsymbol{#1}} 

\newcommand{\nden}{n}
\newcommand{\nddos}{\nden_2}
\newcommand{\nduno}{\nden_1}
\newcommand{\nnum}{N}
\newcommand{\nnmod}{M}

\newcommand{\fdos}{\mu_2}
\newcommand{\funo}{\mu_1}
\newcommand{\conflim}{\bar c}
\newcommand{\conflimmin}{\bar c_\mr{min}}

\begin{document}

\title{Improved Sequential Stopping Rule for\\ Monte Carlo Simulation \thanks{Paper accepted in \emph{IEEE Transactions on Communications.}}}
\author{Luis Mendo and Jos\'e M. Hernando \thanks{E.T.S. Ingenieros de Telecomunicaci\'on, Polytechnic University of Madrid, 28040 Madrid, Spain. E-mail: lmendo@grc.ssr.upm.es.}}
\date{September 2008}

\maketitle

\begin{abstract}
This paper presents an improved result on the negative-binomial Monte Carlo technique analyzed in a previous paper\footnote{L.~Mendo and J.~M. Hernando, ``A simple sequential stopping rule for {Monte Carlo} simulation,'' {\em {IEEE} Trans. Commun.}, vol.~54, no.~2, pp. 231--241, Feb. 2006.} for the estimation of an unknown probability $p$. Specifically, the confidence level associated to a relative interval $[p/\fdos,\, p\funo]$, with $\funo$, $\fdos>1$, is proved to exceed its asymptotic value for a broader range of intervals than that given in the referred paper, and for any value of $p$. This extends the applicability of the estimator, relaxing the conditions that guarantee a given confidence level.

\end{abstract}

\section{Introduction}
\label{parte: intro}

Monte Carlo (MC) methods are widely used for estimating an unknown parameter by means of repeated trials or realizations of a random experiment. An important particular case is that in which the parameter to be estimated is the probability $p$ of a certain event $H$, and realizations are independent. In this setting, the technique of negative-binomial MC (NBMC) \cite{Mendo06} can be used. This technique applies a sequential stopping rule, which consists in carrying out as many realizations as necessary to obtain a given number $\nnum$ of occurrences of $H$. Based on this rule, an estimator is introduced in \cite{Mendo06}, and it is shown to have a number of interesting properties, in the form of respective bounds for its bias, relative precision, and confidence level for a relative interval $[p/\fdos,\, p\funo]$; $\funo$, $\fdos>1$. Specifically, regarding the latter, it is derived in \cite{Mendo06} that the confidence level $c = \Pr[p/\fdos \leq \hat{\va p} \leq p\funo]$ has an asymptotic value $\conflim$ as $p \rightarrow 0$, given by\footnote{
$\gamma(r,x)$ denotes the incomplete gamma function, defined as $\gamma(r,x) = 1/\Gamma(r) \cdot \int_0^x e^{-t}t^{r-1}dt$.}
\begin{equation}
\label{eq: conflim}
\conflim = \gamma(\nnum,(\nnum-1)\fdos)-\gamma \left( \nnum,\frac{\nnum-1}{\funo} \right).
\end{equation}
Furthermore, the confidence level $c$ is assured to exceed $\conflim$ for
\begin{equation}
\label{eq: restr orig en funo y fdos}
\fdos \geq \frac{\nnum+\sqrt{\nnum}}{\nnum-1},\quad
\funo \geq \frac{\nnum-1}{ \nnum - \sqrt{\frac 3 2 \nnum} },
\end{equation}
provided that\footnote{
$\lfloor \cdot \rfloor$ and $\lceil \cdot \rceil$ respectively denote rounding to the nearest integer towards $-\infty$ and towards $\infty$.}
\begin{equation}
\label{eq: restr orig en p}
p < \frac {\nnum-1} {\left\lceil \frac 7 2 \nnum - 1 \right\rceil \funo}.
\end{equation}

In this paper, the sufficient conditions that assure a confidence level $c > \conflim$ for the NBMC estimator are relaxed in two ways:
\begin{itemize}
\item
The restriction on $p$ given by \eqref{eq: restr orig en p} is eliminated, i.e.~$p$ can be an unconstrained value between $0$ and $1$.
\item
The condition for $\funo$ given by \eqref{eq: restr orig en funo y fdos} is weakened, while maintaining the condition for $\fdos$. Thus $\funo$ can be further decreased while having the same guaranteed confidence level given by \eqref{eq: conflim}.
\end{itemize}
The result is presented in Section \ref{parte: B NBMC}, and conclusions are given in Section \ref{parte: concl}.

\section{Result}
\label{parte: B NBMC}

Consider a random experiment, and an event $H$ associated to that experiment (more generally, there may be a set of events associated to the experiment, one of which is of interest). The probability $p$ of event $H$ is to be estimated from independent realizations of the experiment, using the method described in \cite{Mendo06}. Specifically, given\footnote{$\mathbb N$ denotes the set of natural numbers, \{1, 2, \ldots\}.} $\nnum \in \mathbb N$, with $\nnum \geq 3$, realizations are carried out until $\nnum$ occurrences of $H$ are obtained. The number of realizations is thus a negative-binomial random variable\footnote{Random variables are denoted in boldface throughout the paper.} $\va \nden$, from which $p$ is estimated as \cite{Mendo06}
\begin{equation}
\label{eq: B NBMC}
\hat{\va p} = \frac{\nnum-1}{\va \nden}.
\end{equation}

For $\funo$, $\fdos>1$ given, consider the interval $[p/\fdos, p\funo]$, and its associated confidence $c = \Pr[p/\fdos \leq \hat{\va p} \leq p\funo]$. As shown in \cite{Mendo06}, $c$ tends to $\conflim$ given by \eqref{eq: conflim}
as $p \rightarrow 0$.

\begin{proposition}
\label{prop: conf: mayor B NBMC}
For any $p \in (0,\, 1)$, with $\hat{\va p}$ given by \eqref{eq: B NBMC}, the lower bound $c > \conflim$ holds if
\begin{equation}
\label{eq: conf B cond}
\fdos \geq \frac{\nnum+\sqrt{\nnum}}{\nnum-1},\quad
\funo \geq \frac{\nnum-1}{\nnum - \frac 1 2 - \sqrt{\nnum-\frac 1 2}}.
\end{equation}
\end{proposition}

\begin{proof}
Consider $\nnum \geq 3$, $\funo$, $\fdos>1$, and $p \in (0,\, 1)$. Let us define
\begin{align}
\label{eq: nduno B}
\nduno & = \left\lceil \frac{\nnum-1}{p \funo} \right\rceil \\
\label{eq: nddos B}
\nddos & = \left\lfloor \frac{(\nnum-1)\fdos}{p} \right\rfloor.
\end{align}
The confidence $c$ is given by $1-c_1-c_2$
with\footnote{The following notation is used: $k^{(i)} = k(k-1)\cdots(k-i+1)$, $k^{(0)}=1$.}
\begin{equation}
\label{eq: c 1}
\begin{split}
c_1 & = \Pr[\va \nden \leq \nduno-1] \\
& = \frac{p^\nnum}{(\nnum-1)!} \sum_{\nden=\nnum}^{\nduno-1} (\nden -1)^{(\nnum-1)} (1-p)^{\nden -\nnum},
\end{split}
\end{equation}
\begin{equation}
\begin{split}
c_2 & = \Pr[\va \nden \geq \nddos+1] \\
& = \frac{p^\nnum}{(\nnum-1)!} \sum_{\nden =\nddos+1}^{\infty} (\nden -1)^{(\nnum-1)} (1-p)^{\nden -\nnum}.
\end{split}
\end{equation}
Let $\conflim_1$ and $\conflim_2$ be respectively defined as $\lim_{ p \rightarrow 0} c_1$ and $\lim_{ p \rightarrow 0} c_2$. From \cite[appendix C]{Mendo06}, $\conflim_1 = \gamma(\nnum,(\nnum-1)/\funo)$ and $\conflim_2 = 1 - \gamma(\nnum,(\nnum-1)\fdos)$. We will show that $\conflim_1 > c_1$ and $\conflim_2 > c_2$ for $\funo$, $\fdos$ as in \eqref{eq: conf B cond}. This will establish\footnote{It should be noted that, although [1, appendix C] considers $\nddos \geq a$, the actual range of values for $\nddos$ is $\nddos > a-1$. Nonetheless, the proofs in [1, appendix C] can be readily generalized to $\nddos > a-1$.}
that $c > \conflim$.

The inequality $\conflim_2 > c_2$ is equivalent to $\Pr[\va \nden \leq \nddos] > \lim_{p\rightarrow 0} \Pr[\va \nden \leq \nddos]$, which is established in \cite[appendix C]{Mendo06} for $\fdos$ as in \eqref{eq: conf B cond} and $\nddos$ given by \eqref{eq: nddos B}.

In order to show that $\conflim_1 > c_1$, we first note that
\begin{equation}
\begin{split}
\conflim_1 & = \gamma\left( \nnum, \frac{\nnum-1}{\funo} \right) =
\frac{p^\nnum}{(\nnum-1)!} \int_{0}^{ \frac{\nnum-1}{p\funo} } t^{\nnum-1} e^{-pt} dt \\
\label{eq: conflim 1 mayor}
& > \frac{p^\nnum}{(\nnum-1)!} \int_{\nnum-1}^{\nduno-1} t^{\nnum-1} e^{-pt} dt.
\end{split}
\end{equation}
Lemma \ref{lema: 1} given in the Appendix
implies that the right-hand side of \eqref{eq: conflim 1 mayor} will be greater than or equal to that of \eqref{eq: c 1} if $\nduno-1 \leq (\nnum-1/2-\sqrt{\nnum-1/2})/p+1/2$. From \eqref{eq: nduno B}, $\nduno$ is upper bounded by $(\nnum-1)/(p\funo) + 1$. Therefore, in order to assure that $\conflim_1 > c_1$ it is sufficient that
\begin{equation}
\frac{\nnum-1}{p\funo} \leq \frac{\nnum-\frac 1 2 - \sqrt{\nnum-\frac 1 2}} p + \frac 1 2,
\end{equation}
or equivalently
\begin{equation}
\label{eq: funo B p}
\funo \geq \frac{\nnum-1}{\nnum - \frac 1 2 - \sqrt{\nnum - \frac 1 2} + \frac p 2 }.
\end{equation}
Since $p>0$, \eqref{eq: funo B p} holds for $\funo$ as in \eqref{eq: conf B cond}.
\end{proof}

This result removes some of the restrictions that are used in \cite{Mendo06} to assure that $c > \conflim$. Specifically, $p$ can take any value, and the minimum required value for $\funo$ is lower.

\begin{figure}%
\centering%
\includegraphics[width = .8\textwidth]{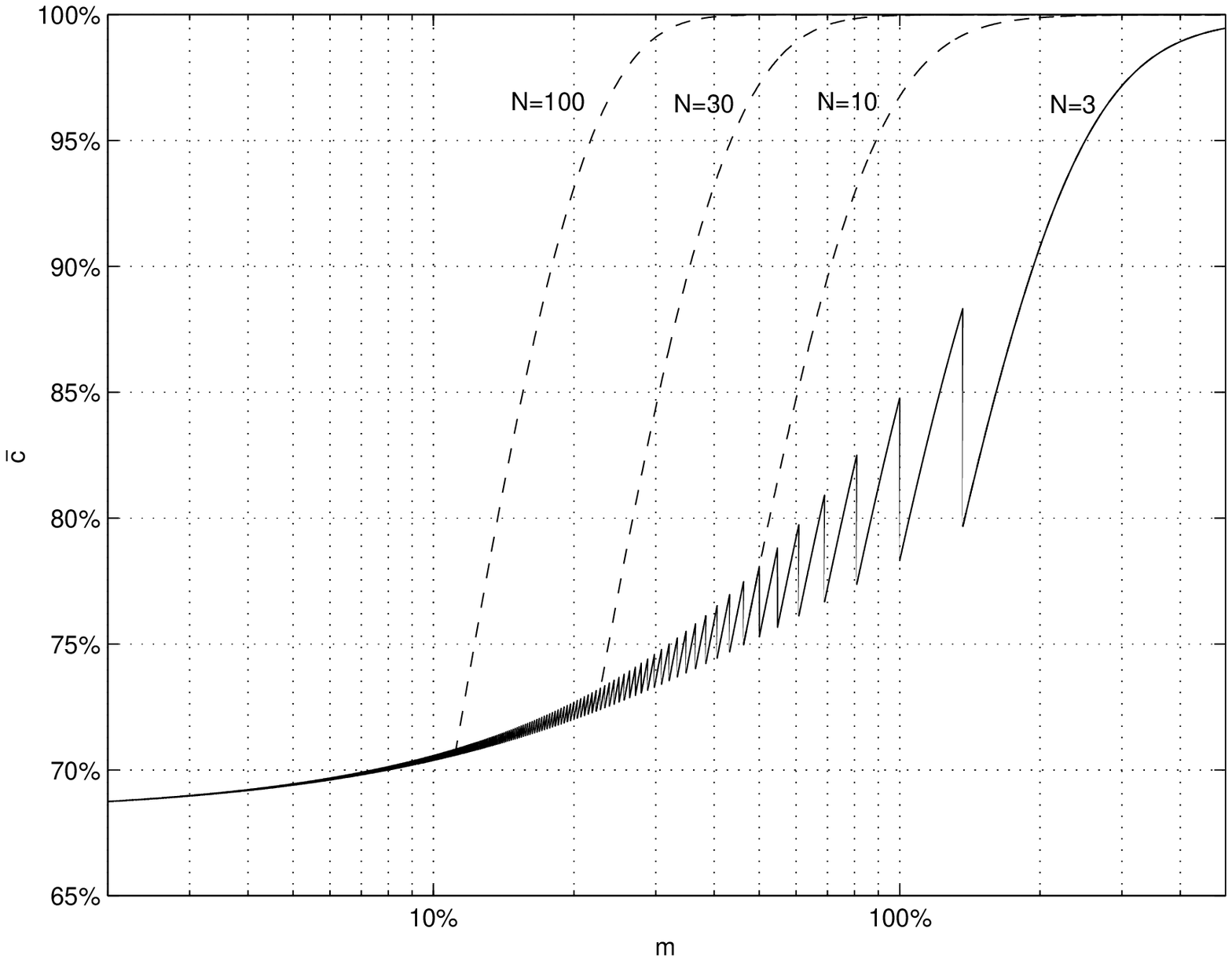}\\%
{\footnotesize -{}-{}- $\conflim$ \qquad --- $\conflimmin$}%
\caption{%
\label{fig: conf-B}%
Guaranteed confidence $\conflim$ and minimum confidence that can be guaranteed $\conflimmin$}%
\end{figure}

For the particular case that $\funo=\fdos=1+m$, where $m>0$ is a relative error margin, it is easily seen that the limiting condition in \eqref{eq: conf B cond} is that for $\fdos$, i.e.
\begin{equation}
\label{eq: m geq}
m \geq \frac{\sqrt\nnum+1}{\nnum-1}.
\end{equation}
The dashed curves in Fig.~\ref{fig: conf-B} depict the guaranteed confidence $\conflim$ (given by \eqref{eq: conflim}) as a function of $\nnum$ and $m$, for $m$ within the allowed range \eqref{eq: m geq}. The solid line represents the minimum confidence $\conflimmin$ that can be guaranteed as a function of $m$. This corresponds to the lowest $\nnum$ permitted by \eqref{eq: m geq} for a given $m$; increasing $\nnum$ gives larger guaranteed confidence levels.

The achievable region in the $(m,\conflim)$ plane is that above the solid curve, in the following sense: for any $(m,\conflim)$ within this region, the confidence level associated to the error margin $m$ can be assured to be greater than $\conflim$, irrespective of $p$; this is accomplished by selecting $\nnum$ according to \eqref{eq: conflim} (or, equivalently, using the curves in \cite[fig. 5(a)]{Mendo06}). Comparing with \cite[fig. 4]{Mendo06}, it is seen that Proposition \ref{prop: conf: mayor B NBMC} enlarges the achievable region, specially for low $m$; besides, it removes the restriction on $p$. Fig.~\ref{fig: conf-B} thus replaces \cite[fig.~4(a)]{Mendo06}; and \cite[figures 4(b) and 5(b)]{Mendo06} are no longer necessary.

As an example, consider the following problem: given an NBMC estimator $\hat {\va p}$ of an unknown $p$ with $\nnum=30$, find the minimum $m$ such that $\Pr[p/(1+m) \leq \hat {\va p} \leq p(1+m)] > 75\%$. From Proposition \ref{prop: conf: mayor B NBMC} and \eqref{eq: conflim}, $m=23.7\%$. The results in \cite{Mendo06}, on the contrary, can only give $m=24.5\%$, since $(23.7\%,75\%)$ is not in the achievable region according to \cite[fig.~4(a)]{Mendo06}; besides, $p<0.224$ is required.

\section{Conclusions}
\label{parte: concl}

In this paper, the statistical characterization of the NBMC estimator introduced in \cite{Mendo06} has been improved by relaxing the conditions that guarantee a certain confidence level. It has been established that, for $p \in (0,\, 1)$ arbitrary, the NBMC estimator has a confidence level better than \eqref{eq: conflim} provided that $\funo$, $\fdos$ satisfy \eqref{eq: conf B cond}. This result extends the range of application of the NBMC estimation technique.

\appendix

\section{Appendix}

The following lemma, used in the proof of proposition \ref{prop: conf: mayor B NBMC}, is now established.

\begin{lemma}
\label{lema: 1}

Given $\nnum$, $\nden^\ast \in \mathbb N$ with $\nnum \geq 3$ and
\begin{equation}
\label{eq: lema 1 cond}
\nden^\ast \leq \frac{\nnum - \frac 1 2 - \sqrt{\nnum - \frac 1 2}} p + \frac 1 2
\end{equation}
the following inequality holds:
\begin{equation}
\label{eq: desig lema 1}
\int_{\nnum-1}^{\nden^\ast} \nden^{\nnum-1} e^{-p\nden} d\nden \geq \sum_{n=\nnum}^{\nden^\ast} (n-1)^{(\nnum-1)} (1-p)^{n-\nnum}.
\end{equation}
\end{lemma}

\begin{proof}
We first note that the sub-integral function is increasing for $\nden <(\nnum-1)/p$, and is convex for $(\nnum-1-\sqrt{\nnum-1})/p < \nden <(\nnum-1)/p$. As figure \ref{fig: dem 1} illustrates,
\begin{figure}%
\centering%
\includegraphics[clip, width = .8\textwidth]{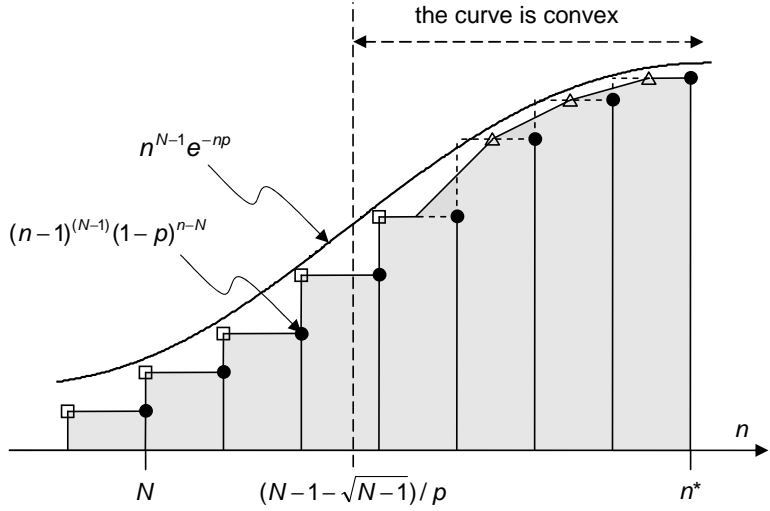}\\%
\caption{%
\label{fig: dem 1}%
Graphical representation for the proof of Lemma \ref{lema: 1}}%
\end{figure}%
each term of the sum in \eqref{eq: desig lema 1} can be identified with the area of a rectangle of unit width. Specifically, the term corresponding to a given $\nden$ is associated to the rectangle that extends horizontally from $\nden-1$ to $\nden$ in the figure. In addition, for the rectangles situated to the right of $(\nnum-1-\sqrt{\nnum-1})/p$ (where the sub-integral function is convex), the flat tops are replaced by straight lines joining the centers, without altering the total area. The inequality \eqref{eq: desig lema 1} will hold if the area below the curve in the interval $(\nnum-1,\, \nden^\ast)$ is larger than the shaded area in the figure. We divide this interval in two: $(\nnum-1,\, (\nnum-1-\sqrt{\nnum-1})/p)$ and $((\nnum-1-\sqrt{\nnum-1})/p,\, \nden^\ast)$, and require that in each of these intervals the area of the curve be larger than the part of the shaded area corresponding to that interval. In the first interval, since the curve is increasing, it suffices that the curve be above the square marks for $\nden \leq \lfloor (\nnum -1-\sqrt{\nnum-1})/p \rfloor$, as shown in the figure. In the second interval, since the curve is increasing and convex, it suffices that the curve be above the square mark located at $\lfloor (\nnum -1-\sqrt{\nnum-1})/p \rfloor +1$ and above the triangle marks. As a result, to establish \eqref{eq: desig lema 1} it is sufficient that
\begin{itemize}
\item[(i)]
the sub-integral function be above the square marks for $\nnum \leq \nden \leq (\nnum-1-\sqrt{\nnum-1})/p+1$; and
\item[(ii)]
the sub-integral function be above the triangle marks for $(\nnum-1-\sqrt{\nnum-1})/p+1 < \nden < \nden^\ast$.
\end{itemize}
We analyze these conditions separately.

(i) With $x$ defined as
\begin{equation}
\begin{split}
x & = \frac 1 p \ln \frac{(\nden-1)^{\nnum-1} e^{-(\nden-1)p}} {(\nden -1)^{(\nnum-1)} (1-p)^{\nden -\nnum}} \\
& = -\frac 1 p \sum_{i=1}^{\nnum-1} \ln\left( 1-\frac {i-1} {\nden-1} \right) - \frac{\nden-\nnum}{p} \ln(1-p) - (\nden-1),
\end{split}
\end{equation}
condition (i) is expressed as $x \geq 0$ for $\nden \leq (\nnum-1-\sqrt{\nnum-1})/p+1$. Defining $\nnmod=\nnum-1$ and $\nu = (\nden-1)p$,
\begin{equation}
\label{eq: x}
x = -\frac 1 p \sum_{i=1}^{\nnum-1} \ln\left( 1-\frac {(i-1)p} {\nu} \right) - \frac 1 p \left(\frac{\nu}{p}-\nnmod \right) \ln(1-p) - \frac {\nu}{p}.
\end{equation}
Using the Taylor expansion $\ln(1-t) = -\sum_{j=1}^\infty t^j/j$, $|t|<1$, \eqref{eq: x} is transformed into $x = \sum_{j=0}^\infty x_j p^j$ with
\begin{equation}
\label{eq: x j}
x_j = \frac 1 {(j+1)\nu^{j+1}} \sum_{i=1}^{\nnum-1} (i-1)^{j+1} + \frac{\nu}{j+2}-\frac{\nnmod}{j+1}.
\end{equation}
The term $x_0$ is easily seen to be nonnegative for $\nu \leq \nnmod - \sqrt{\nnmod}$, i.e.~for $\nden \leq (\nnum-1-\sqrt{\nnum-1})/p+1$. We now prove that the remaining coefficients $x_j$, $j \geq 1$ are also nonnegative for $\nu \leq \nnmod-\sqrt{\nnmod}$.

We begin with the case $\nnum \geq 5$, $j \geq 2$. Using the inequality
\begin{equation}
\begin{split}
\sum_{i=1}^{\nnum-1} (i-1)^{j+1} & > \int_1^{\nnum-1} (i-1)^{j+1}di \\
& = \frac{(\nnum-2)^{j+2}}{j+2}
= \frac{(\nnmod-1)^{j+2}}{j+2}
\end{split}
\end{equation}
in \eqref{eq: x j}, we can bound
\begin{equation}
\label{eq: x j cota}
x_j > \frac {(\nnmod-1)^{j+2} + (j+1)\nu^{j+2} - (j+2) \nnmod \nu^{j+1}}
{(j+1)(j+2)\nu^{j+1}}.
\end{equation}

The denominator in \eqref{eq: x j cota} is positive. Let $y_j$ denote the numerator. We compute
\begin{equation}
\frac{\partial y_j}{\partial \nu} = (j+2)(j+1) (\nu-\nnmod) \nu^j <0,
\end{equation}
from which it suffices to consider $\nu=\nnmod-\sqrt{\nnmod}$. Expressing
\begin{equation}
\begin{split}
y_j|_{\nu=\nnmod-\sqrt{\nnmod}} & = {\nnmod}^{j+2} \left( 1- \frac{1}{\sqrt{\nnmod}}\right)^{j+1} \\
& \quad \cdot \left[ \left(1-\frac{1}{\sqrt{\nnmod}}\right) \left(1+\frac{1}{\sqrt{\nnmod}}\right)^{j+2} \right. \\
& \quad + \left. (j+1) \left(1-\frac{1}{\sqrt{\nnmod}}\right) - (j+2) \right]
\end{split}
\end{equation}
and denoting the bracketed term by $Y_j$, we now compute the following partial derivatives as if $j$ were a continuous variable:
\begin{equation}
\begin{split}
\label{eq: djdnu Yj}
\frac{\partial Y_j}{\partial j} & = \left(1-\frac{1}{\sqrt{\nnmod}}\right) \left(1+\frac{1}{\sqrt{\nnmod}}\right)^{j+2} \\
& \quad \cdot \ln \left(1+\frac{1}{\sqrt{\nnmod}}\right) -\frac{1}{\sqrt{\nnmod}}
\end{split}
\end{equation}
\begin{equation}
\label{eq: d2jdnu Yj}
\frac{\partial^2 Y_j}{\partial j^2} = \left(1-\frac{1}{\sqrt{\nnmod}}\right) \left(1+\frac{1}{\sqrt{\nnmod}}\right)^{j+2} \ln^2 \left(1+\frac{1}{\sqrt{\nnmod}}\right).
\end{equation}
The right-hand side of \eqref{eq: d2jdnu Yj} is positive, and using the inequality $\ln(1+t)>t-t^2/2$ we can bound \eqref{eq: djdnu Yj} for $j=2$ as
\begin{equation}
\begin{split}
\left. \frac{\partial Y_j}{\partial j} \right|_{j=2} & = \left(1-\frac{1}{\sqrt{\nnmod}}\right) \left(1+\frac{1}{\sqrt{\nnmod}}\right)^4 \frac{1}{\sqrt{\nnmod}} \\
& \quad \cdot \left(1-\frac{1}{2\sqrt{\nnmod}}\right) -\frac{1}{\sqrt{\nnmod}} > 0.
\end{split}
\end{equation}
Therefore the right-hand side of \eqref{eq: djdnu Yj} is positive for $j \geq 2$. Consequently, in order to establish that $y_j \geq 0$, it suffices to show that $Y_2 \geq 0$. Defining $m=1/\sqrt{\nnmod}$, $Y_2$ can be expressed as $-m^2 (m^3+3m^2+2m-2)$. According to Descartes' sign rule, this polynomial has only one positive root. For $m \rightarrow \infty$ the polynomial takes negative values, and for $m=1/2$ it is positive. Therefore, it is positive for $m \leq 1/2$, i.e.~for $M \geq 4$, or $\nnum \geq 5$.

For $\nnum = 4$, $j \geq 2$, we have
\begin{equation}
\label{eq: x j nnum 4}
x_j = \frac {(j+2)(1+2^{j+1}) + (j+1) \nu^{j+2} -3(j+2) \nu^{j+1}}
{(j+2)(j+1)\nu^{j+1}}
\end{equation}
Let $z_j$ denote the numerator in \eqref{eq: x j nnum 4}. Since
\begin{equation}
\frac{\partial z_j}{\partial \nu} = (j+2)(j+1)\nu^j (\nu-3) < 0,
\end{equation}
it suffices to consider $\nu=3-\sqrt{3}$. Bounding $z_j$ as
\begin{equation}
z_j > (j+2)(2^{j+1}-3\nu^{j+1}) + (j+1) \nu^{j+2},
\end{equation}
$z_j|_{\nu=3-\sqrt{3}}$ is seen to be positive for $j \geq 2$.

For $\nnum = 3$, $j \geq 2$,
\begin{equation}
x_j = \frac {j+2 + (j+1) \nu^{j+2} -2(j+2) \nu^{j+1}}
{(j+2)(j+1)\nu^{j+1}},
\end{equation}
and similar arguments to those for $\nnum = 4$, $j \geq 2$ show that $x_j > 0$.

For $\nnum \geq 3$, $j=1$, using the identity
\begin{equation}
\begin{split}
\sum_{i=1}^{\nnum-1} (i-1)^2 & = \frac{(\nnum-2)^3} 3 + \frac{(\nnum-2)^2} 2 + \frac{\nnum-2} 6 \\
& = \frac{(\nnmod-1)^3} 3 + \frac{(\nnmod-1)^2} 2 + \frac{\nnmod-1} 6
\end{split}
\end{equation}
we obtain from \eqref{eq: x j}
\begin{equation}
\label{eq: x j j 1}
x_j = \frac{4\nu^3-6\nnmod\nu^2 + 2(\nnmod-1)^3+3(\nnmod-1)^2+\nnmod-1}{12\nu^2}.
\end{equation}
From Descartes' sign rule, the numerator of \eqref{eq: x j j 1} considered as a polynomial in $\nu$ has two positive roots at most. This polynomial is positive for $\nu=0$ and for $\nu \rightarrow \infty$, and negative for $\nu=\nnmod$. Thus, it will be positive for $\nu \leq \nnmod-\sqrt{\nnmod}$ if it is for $\nu = \nnmod-\sqrt{\nnmod}$. Substituting this $\nu$ and defining $m=\sqrt{\nnmod}$, the numerator is expressed as $m^2(3m^2-4m+1)$, which is positive for $m>1$, or equivalently $M>1$, and thus for $\nnum \geq 3$.

(ii) With $x'$ defined as
\begin{equation}
\begin{split}
x' & = \frac 1 p \ln \frac{\left(\nden-\frac 1 2 \right)^{\nnum-1} e^{-\left(\nden-\frac 1 2 \right)p}} {(\nden -1)^{(\nnum-1)} (1-p)^{\nden -\nnum}} \\
& = -\frac 1 p \sum_{i=1}^{\nnum-1} \ln\left( 1-\frac {i-\frac 1 2} {\nden-\frac 1 2} \right) \\
& \quad - \frac{\nden-\nnum}{p} \ln(1-p) - \left(\nden-\frac 1 2 \right),
\end{split}
\end{equation}
and taking into account \eqref{eq: lema 1 cond}, in order to fulfil condition (ii) it is sufficient that $x' \geq 0$ for $\nden \leq (\nnum-1/2-\sqrt{\nnum-1/2})/p+1/2$. Defining $\nnmod'=\nnum-1/2$ and  $\nu' = (\nden-1/2)p$,
\begin{equation}
\begin{split}
x' & = -\frac 1 p \sum_{i=1}^{\nnum-1} \ln\left( 1-\frac {\left(i-\frac 1 2 \right)p} {\nu'} \right) \\
& \quad - \frac 1 p \left( \frac{\nu'}{p} - \nnmod' \right) \ln(1-p) - \frac{\nu'}{p}.
\end{split}
\end{equation}
Proceeding as with $x$, we can express $x'=\sum_{j=0}^\infty x'_j p^j$ with
\begin{equation}
\label{eq: x prima j lema 1}
x'_j = \frac 1 {(j+1)\nu'^{j+1}} \sum_{i=1}^{\nnum-1} \left(i-\frac 1 2 \right)^{j+1} + \frac{\nu'}{j+2}-\frac{\nnmod'}{j+1}.
\end{equation}
$x'_0$ is seen to be nonnegative for $\nu' \leq \nnmod' - \sqrt{\nnmod'-1/4}$, and thus for $\nu' \leq \nnmod' - \sqrt{\nnmod'}$, i.e.~for $\nden \leq (\nnum-1/2-\sqrt{\nnum-1/2})/p+1/2$. We now prove that the remaining coefficients $x'_j$, $j \geq 1$ are also nonnegative for $\nu' \leq \nnmod' - \sqrt{\nnmod'}$.

We begin with the case $\nnum \geq 5$, $j \geq 2$. Using the inequality
\begin{equation}
\begin{split}
\sum_{i=1}^{\nnum-1} \left( i-\frac 1 2 \right)^{j+1} & > \int_{1/2}^{\nnum-1} \left( i - \frac 1 2 \right)^{j+1} di \\
& = \frac{\left(\nnum-\frac 3 2 \right)^{j+2}}{j+2} = \frac{\left(\nnmod'-1 \right)^{j+2}}{j+2}
\end{split}
\end{equation}
in \eqref{eq: x prima j lema 1}, we can bound
\begin{equation}
\label{eq: y j cota}
x'_j > \frac {\left(\nnmod'-1 \right)^{j+2} + (j+1)\nu'^{j+2} - (j+2) \nnmod' \nu'^{j+1}} {(j+1)(j+2)\nu'^{j+1}}.
\end{equation}
The denominator in \eqref{eq: y j cota} is positive, and the numerator is as that in \eqref{eq: x j cota} with $\nnmod$ replaced by $\nnmod'$ and $\nu$ replaced by $\nu'$. It stems that $x'_j>0$ for $\nnmod' \geq 4$, i.e.~$\nnum \geq 5$, and $j \geq 2$.

For $\nnum=4$, $j \geq 2$, \eqref{eq: x prima j lema 1} gives
\begin{equation}
\begin{split}
x'_j & = \Big[ (j+2) \left(\left (\textstyle \frac 1 2 \right)^{j+1}+ \left(\textstyle \frac 3 2 \right)^{j+1}+ \left(\textstyle \frac 5 2 \right)^{j+1} \right) + (j+1) \nu'^{j+2} \\
& \quad - \textstyle \frac 7 2 (j+2) \nu'^{j+1} \Big] \Big/ \left[ (j+2)(j+1)\nu'^{j+1} \right],
\end{split}
\end{equation}
which is shown to be positive with analogous arguments as for $x_j$.

For $\nnum=3$, $j \geq 2$,
\begin{equation}
\begin{split}
x'_j & = \Big[ (j+2) \left( \left(\textstyle \frac 1 2 \right)^{j+1}+ \left(\textstyle \frac 3 2 \right)^{j+1} \right) + (j+1) \nu'^{j+2} \\
& \quad - \textstyle \frac 5 2 (j+2) \nu'^{j+1} \Big] \Big/
\left[ (j+2)(j+1)\nu'^{j+1} \right],
\end{split}
\end{equation}
and similarly it is shown to be positive.

For $\nnum \geq 3$, $j=1$, using the identity
\begin{equation}
\begin{split}
\sum_{i=1}^{\nnum-1} \left(i - \frac 1 2\right)^2 & = \frac{\left(\nnum- \frac 3 2 \right)^3} 3 + \frac{\left(\nnum- \frac 3 2 \right)^2} 2 + \frac{\nnum- \frac 3 2} 6 \\
& = \frac{\left(\nnmod'-1 \right)^3} 3 + \frac{\left(\nnmod'-1 \right)^2} 2 + \frac{\nnmod'-1} 6
\end{split}
\end{equation}
we obtain an expression for $x'_j$ which coincides with \eqref{eq: x j j 1} replacing $\nnmod$ by $\nnmod'$ and $\nu$ by $\nu'$. Therefore, $x'_j$ is positive for $\nnmod'>1$, and thus for $\nnum \geq 3$.

According to the foregoing, conditions (i) and (ii) hold for $\nnum \geq 3$. This establishes the stated result \eqref{eq: desig lema 1}.
\end{proof}

\section*{Acknowledgment}

The authors would like to thank the anonymous reviewers and the Editor for Wireless Systems Performance, F. Santucci, for their helpful comments.


\begin{thebibliography}{1}

\bibitem{Mendo06}
L.~Mendo and J.~M. Hernando, ``A simple sequential stopping rule for {Monte
  Carlo} simulation,'' {\em {IEEE} Trans. Commun.}, vol.~54, no.~2,
  pp. 231--241, Feb. 2006.

\end{thebibliography}
\end{document}